\documentclass[preprint,proceedings]{rmaa}

\suppressfulladdresses 

\usepackage{paralist}

\usepackage{psfrag,color}

\newcommand{\hb}{H$\beta$}
\newcommand{\feii}{FeII}
\newcommand{\hbbc}{H$\beta$$_{BC}$}
\newcommand{\oiii}{[OIII]}
\newcommand{\gs}{$\Gamma$$_{soft}$}
\newcommand{\civ}{CIV}
\newcommand{\rfe}{R$_{FeII}$}

\SetYear{2007} \SetConfTitle{The Nuclear Region, Host Galaxy and
Environment of Active Galaxies}

\title{Our Search for an H-R Diagram of Quasars}

\author{Jack W. Sulentic\altaffilmark{1},
  Sebastian Zamfir\altaffilmark{1},
  Paola Marziani\altaffilmark{2} and
  Deborah Dultzin\altaffilmark{3}}
\altaffiltext{1}{Department of Physics \& Astronomy, University of
Alabama, Tuscaloosa, 35487 USA; giacomo@merlot.astr.ua.edu.}

\altaffiltext{2}{Osservatorio Astronomico di Padova, 35122 Padova,
Italia.}

\altaffiltext{3}{Instituto de Astronomia, UNAM, M\'exico.}

\shortauthor{Sulentic et al.}
\shorttitle{RevMexAA(SC) Our Search for an H-R Diagram of Quasars}

\listofauthors{J. W. Sulentic, S. Zamfir, P. Marziani, \& D.
Dultzin}
\indexauthor{Sulentic, J. W.} \indexauthor{Zamfir, S.}
\indexauthor{Marziani, P.} \indexauthor{Dultzin, D.}

\abstract{The 4D Eigenvector 1 parameter space was introduced seven
  years ago as an attempt at multiwavelength spectroscopic
  representation of quasars. It appears to be the most effective
  diagnostic space for unifying the diversity of broad line AGN. This
  progress report shows that the diagnostic power of 4DE1 is confirmed
  using optical spectra from the SDSS, UV spectra from HST and X-ray
  spectra from XMM. Our introduction of the population A-B concept
  continues to provide useful insights into quasar diversity. Largely
  radio-quiet, high accreting, low BH mass Pop. A sources (FWHM
  H$\beta$$\leq$ 4000 km s$^{-1}$) show strong FeII emission, a soft X-ray
  excess and a CIV profile blueshift. Low accreting large BH mass
  Pop. B quasars (FWHM H$\beta$$>$ 4000 km s$^{-1}$) include most radio-loud AGN
  and show weak FeII emission and little evidence for a soft X-ray
  excess or a CIV blueshift.}

\resumen{El espacio de par\'{a}metros conocido como 4DE1
(Eigenvector 1 en cuatro dimenciones) fue introducido hace siete
a\~{n}os en un intento de representar las propiedades
espectrosc\'{o}picas en multifrecuencias de los cuasares. Este
espacio de diagn\'{o}stico ha demoastrado ser el m\'{a}s adecuado
para intentar unificar la diversidad de AGNs con l\'{\i}neas anchas.
Aqui reportamos que el poder de diagn\'{o}stico del espacio de
par\'{a}metros 4DE1 se confirma usando espectros \'{o}pticos tomados
del SDSS, espectros UV del HST y datos de rayos X del satelite XMM.
La introducci\'{o}n del concepto de Poblaci\'{o}nes A y B, continua
brindandonos una manera \'{u}til de entender la diversidad de los
cuasares. La gran mayoria de los objetos radio callados son de
poblaci\'{o}n A (FWHM H$\beta$$\leq$ 4000 km s$^{-1}$) y tienen
altas tasas de acrecion y relativamente bajas masas del Hoyo Negro
nuclear. Son potentes emisores de FeII, del llamado exceso en rayos
X suaves, y muestran corrimientos al azul en los perfiles de las
lineas de CIV. Los objetos de poblaci\'{o}n B (FWHM H$\beta$$>$ 4000
km s$^{-1}$) incluyen a la mayoria de los objetos radio fuertes, con
mayores masas del Hoyo Negro y bajas tasas de acrecion. Estos
muestran debil (o nula) emisi\'{o}n de FeII, del exceso en rayos X
suaves y de corrimientos al azul de las lineas de CIV. }

\addkeyword{galaxies: active}
\addkeyword{galaxies:quasars: emission
lines}
\addkeyword{galaxies:quasars: general}

\begin{document}
\maketitle

\section{Introduction}
\label{sec:intro}

A NED search for references to one of the hundred brightest AGN
(e.g. PG quasars; \citealt{BG92}, hereafter BG92) will frequently
yield $\sim$50-200 ``hits'', but among these one will often find
only very few papers dealing with optical/UV spectroscopy of that
source. This is surprising especially if one considers that such
spectra offer the most direct insights into the geometry and
kinematics of the central broad and narrow line regions.
Multiwavelength astronomy is apparently still in its infancy with
most papers dealing with monochromatic measures. While AGN
unification schemes are popular--up till recently--no spectroscopic
unification has existed for AGN. This is due to at least two things:
1) until relatively recently few sources with moderate to high
resolution and S/N spectroscopy existed for AGN and 2) there has
been a common belief that AGN spectra are basically the same. The
SDSS has ameliorated the former problem, although caution is
required if one considers spectra for sources fainter than
g$\sim$17.5.

We have been searching for a spectroscopic unification embracing all
broad line AGN for the past 10+ years. Our hope was to find a
diagnostic space that could serve somewhat the same role as the H-R
Diagram serves for stellar studies. The H-R Diagram functions well
in 2D although its full power requires exploitation of more
parameter dimensions. Certainly an equivalent spectroscopic
diagnostic space for quasars will require more than two dimensions
if only to remove the degeneracy between source physics and
line-of-sight orientation -- as drivers of measured source
properties. This is a problem that does not afflict the stellar H-R
diagram. In 2000 we proposed a 4D Eigenvector 1 parameters space
\citep[][; hereafter 4DE1]{Sulentic00a,Sulentic00b} as the optimal
vehicle for emphasizing spectroscopic diversity while at the same
time contextualizing the diverse types of broad-line emitting
sources.

Our 4DE1 parameter space  provides a potentially fundamental
discrimination between major AGN classes. 4DE1 space
\citep{Sulentic00a,Sulentic00b} incorporates all  of the
statistically significant line profile similarities and differences
that are currently known. 4DE1 has roots in the Principal Component
Analysis (PCA) of the Bright Quasar Sample (Eigenvector 1; BG92) as
well as in correlations that emerged from ROSAT
\citep[e.g.][]{Wang96}. 4DE1 as we define it involves BG92 measures:
1) full width half maximum of broad \hb\ (FWHM \hb) and 2)
equivalent width ratio of optical \feii\ and broad \hb: R$_{FeII}$ =
W(\feii)/W(\hbbc). We added 3) a measure of the soft X-ray photon
index (\gs) and  4) a measure of \civ\ broad line profile
displacement at half maximum (c(1/2)) to arrive at our 4DE1 space.
Other points of departure from BG92 involve explicit comparison of
RQ and RL sources as well as subordination of \oiii\ measures
(although see \citealt{Zamanov02,Marziani03b}). Finally we divide
AGN into two populations using a simple division at FWHM \hbbc\
=4000 km s$^{-1}$ with sources narrower and broader than this value
designated populations A and B, respectively. The latter distinction
was motivated by the observation that almost all RL sources show
FWHM \hbbc\ $>$4000 km s$^{-1}$ \citep{Sulentic00b}. This
distinction turns out to be more effective for highlighting
spectroscopic differences than the more traditional divisions into:
1) RQ-RL sources  and 2) NLSy1-BLSy1 sources defined with FWHM
\hbbc\ $<$ and $>$ 2000 km s$^{-1}$, respectively.

Adopted 4DE1 key parameters were chosen with the following two
requirements in mind: 1) measures that showed large intrinsic
variance ($\Sigma$) and 2) measures that could be made with high
precision ($\sigma$). Many/most other spectroscopic measures show
correlation in 4DE1 and/or pop.A-B dichotomy (e.g.optical Balmer
line asymmetry, UV SIII]1892/CIII]1909 ratio, hard X-ray spectral
index) but do not show large enough variance, or cannot be obtained
for large enough numbers of sources with suitable precision, to
permit adoption as key parameters (e.g. $\Sigma$$>$20$\sigma$). All
line measures involve suitably corrected broad line components
(reduction procedures are described in
\citealt{Marziani96,Marziani03a,Sulentic07}). In simplest terms the
key parameters can be said to measure: 1) the dispersion in (low
ionization line) BLR cloud velocities, 2) the relative strengths of
FeII and \hb\ emission, 3) the strength of a soft X-ray excess and
4) the amplitude of systematic radial motions in (high ionization
line) BLR emitting gas. In less simple terms (i.e. more model
dependent) we likely have: 1) three or four orthogonal variables
sensitive to the Eddington ratio, 2) two or three variables
sensitive to source inclination, as well as 3) one variable
sensitive to black hole mass (FWHM) and another to nebular physics
(n$_{e}$: \rfe). Taken together they offer the most direct clues
about the geometry and kinematics of the broad line region (BLR).

Source occupation in principal planes of 4DE1 can be found in
\citet{Sulentic00a,Sulentic00b,Sulentic07}. These studies have made
use of ground based optical spectra for the H$\beta$ region, HST
archival FOS/STIS UV spectra for the region of CIV and ROSAT X-ray
spectra. We have recently (Zamfir et al. 2008 in preparation)
exploited the SDSS database which offers major advantages:  1)
uniform, high resolution ($\sim$1$\AA$) and high S/N spectra for the
300+ brightest quasars and 2) broad wavelength coverage
(3000-9000$\AA$). Our adopted redshift limit z=0.7 allows measures
of H$\beta$ and neighboring lines. This has greatly expanded the
number of bright quasars in the magnitude and color range of the PG.
It also provides a sample with the proper ratio of RL to RQ sources.
Uniform FIRST radio measures allow us to explore the differences
between RQ and RL sources with a highly complete ($\sim$80\%; see
\citealt{Jiang07}) sample. Figure 1 offers an introduction to the
optical plane of 4DE1 in the light of the SDSS era and it is
reassuring to note that source occupation confirms our previous
work. Similarly we describe preliminary X-ray results that come from
the growing XMM-NEWTON spectroscopic sample. We attempt to summarize
here the many insights that 4DE1 offers as a diagnostic space that
offers a clearer context in which to interpret individual sources.

\begin{figure*}[!t]
  \resizebox{12.5cm}{!}{\rotatebox{0}{\includegraphics{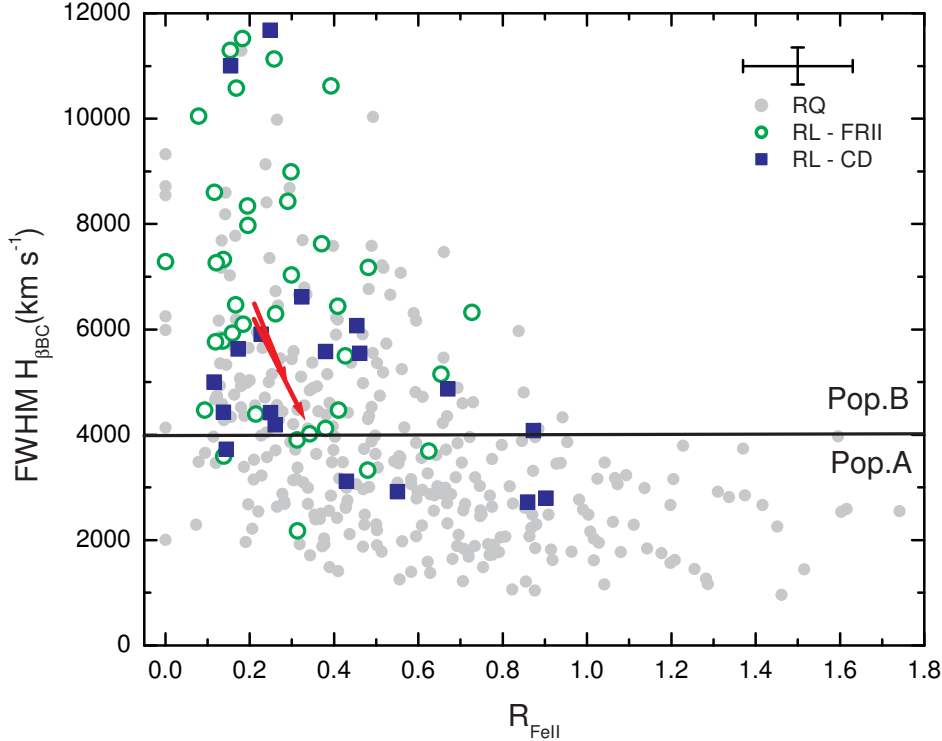}}}
  \caption{The SDSS bright quasar sample in the optical plane of 4DE1 space.
   }
  \label{fig:simple}
\end{figure*}

1) 4DE1 makes clear that broad line emitting AGN show a wide range
of spectroscopic properties -- they are not all alike -- the
intrinsic dispersion in 4DE1 parameters is much larger than
measurement errors. The cross in the upper right corner of Figure 1
indicates typical (median) measurement uncertainties which
correspond roughly to 3$\sigma$. FWHM H$\beta$ measures range from
1000 (smallest known $\sim$ 630 km s$^{-1}$; \citealt{Grupe99}) to
$\sim$23000 km s$^{-1}$ in our sample (largest known $\sim$40000 km
s$^{-1}$; \citealt{Wang05}, in H$\alpha$) and \rfe measures from 0
to $\sim$1.8 (most extreme known \rfe$\sim$6; \citealt{Lipari93}).
Note that SDSS calls sources with FWHM H$\beta$$<$1000 km s$^{-1}$
galaxies rather than quasars and FeII emitters with RFE$>$1.8 were
rarely found in the SDSS quasar catalog. They are extremely red
objects (strong IRAS sources as well; see
\citealt{Lipari93,Veron06}) and they require a special analysis of
the FeII emission. These objects are therefore not included in our
sample. We regard detection of (broad) FeII emission as a
requirement for Type 1 quasar status. Our previous work involved
many more sources with \rfe upper limits reflecting the sensitivity
of FeII detection to continuum s/n especially for broader line
sources \citep[see][Figure 3]{Marziani03a}. SDSS spectra for bright
quasars  reveal less than 1\% of broad-line emitting sources with
undetected FeII emission suggesting that FeII emission is an
ubiquitous property of Type 1 AGN.

Figure 1 tells us that average quasar spectra that do not allow for
the wide observed parameter dispersion are somewhat like average
stellar spectra involving the entire OBAFGKM sequence
\citep[see][;Zamfir et al. 2008 in
preparation]{Sulentic02,Sulentic07,Bachev04} for average spectra in
the 4DE1 context).

2) Earlier studies \citep{Sulentic00a,Sulentic00b} suggested that
the RQ majority of AGN showed a larger 4DE1 domain space occupation
region than RL sources and this is well confirmed with the SDSS
sample. RQ sources (grey filled symbols in Figure 1) form what can
be described as a correlation ``main sequence''. There is a large
parameter domain where no sources are found. In other words we see a
large dispersion in parameter measures but a restricted dispersion
in combinations of these measures (e.g. no broad Balmer line
profiles [FWHM H$\beta$$>$ 8000 km s$^{-1}$] with strong FeII
emission [RFE$>$0.4]). Narrow Line Seyfert 1 sources (NLSy1) do not
emerge as a distinct class but rather as the lower edge of the RQ
sequence. If there is a division or dichotomy in source properties
it occurs near FWHM H$\beta$=4000 km s$^{-1}$.

3) RL sources (blue filled squares/green open circles) occupy one
end of the RQ sequence and overlap with about 25\% of RQ sources.
This domain space separation is highly significant
(P$\sim$3$\times$10$^{-7}$ that RQ and RL occupy the same domain;
Zamfir et al. 2008) and credible because we are dealing with
unusually complete/balanced RQ and RL samples. These are therefore
the only RQ sources that are spectroscopically indistinguishable
from RL AGN. We define a source as RL if logL$_{20cm}(erg s^{-1}
Hz^{-1})$=31.5 (this is approximately a radio/optical flux density
ratio R$_{K}$=70; \citealt{Kellermann89}). This corresponds to the
radio properties of the weakest observed double-lobed (assumed
unbeamed in an orientation unification scenario) source in our old
and new samples \citep[][Zamfir et al. 2008]{Sulentic03}. The
probability of radio loudness in Figure 1 increases from P$\sim$0.01
at the extreme RQ end of the source sequence (where NLSy1 are found;
\citealt{Komossa06} estimates P$\sim$0.025 for NLSy1) to
P$\sim$0.15-0.18 at the opposite end where the vast majority of RL
sources are found.

4) Few RL sources are found below FWHM \hbbc\ =4000 km s$^{-1}$.
Those that do likely involve sources where the putative accretion
disk is seen face-on (and radio jets pole-on) \citep[][Zamfir et al.
2008]{Sulentic03}. We can infer this because we see a separation
between the majority of core-dominated sources (CD--filled squares)
and lobe-dominated (LD--open circles) emission. Most RL sources with
large and small FWHM H$\beta$ values are lobe- and core-dominated
(alternatively steep and flat radio spectrum sources) respectively
again as expected from simple orientation unification schemes (e.g.
\citealt{UP95, OB82}) where Balmer emission arises in a flattened
cloud distribution (usually called an accretion disk) and the
radio-jets are aligned perpendicular to the cloud distribution. Red
arrows in Figure 1 indicate the median displacement in 4DE1 optical
plane coordinates between FRII and CD sources in the old and new
source samples \citep[][Zamfir et al. 2008]{Sulentic03}. This is
consistent with LD sources viewed as the (misaligned) parent
population of RL sources and with CD sources as the aligned (near
pole-on and Doppler boosted) sources. Some CD (largest FWHM) and LD
(smallest FWHM) sources do not follow the clear trend indicated by
the red arrows in Figure 1. These likely represent the (about
10-15\% in our RL sample) ``misaligned'' sources where the radio
jets are far from perpendicular to the accretion disk plane.

5) All reasonably resolved LD sources in our SDSS sample show FRII
morphology suggesting that FRI sources are rare among broad-line RL
emitters. Some weak FRI sources may be present among the RQ sources
but they show extreme core/lobe ratios and are effectively filtered
out of the FIRST survey (dynamic range limits and spatial frequency
attenuation). We find no evidence for a significant
radio-intermediate population bridging the RQ and RL sources. While
weak radio jets are observed in some RQ quasars
\citep[e.g.][]{Ulvestad05,Leipski06}, the majority show radio
emission dominated by star formation in the host galaxies (e.g. PG
quasars follow the radio-FIR correlation; see \citealt{Haas03}).
There is no evidence for a continuum of radio jet properties from
largely LD RL to largely CD RQ sources although a few RQ show
unusually strong radio jet-lobe structures \citep[e.g.][]{G-B02}.
The strongest support for this conclusion comes from 4DE1 (Figure 1)
where RL sources show a preferred domain space occupation relative
to the RQ majority.

\begin{figure}
  \includegraphics[width=\columnwidth]{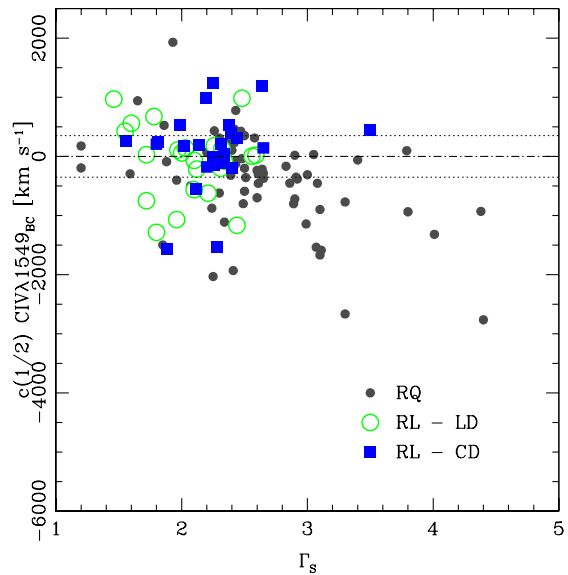}
  \caption{Our bright quasar sample in a UV -- X-ray plane of 4DE1 space.
  Dashed lines are 2$\sigma$ confidence intervals for zero CIV shift.}
  \label{fig:simple}
\end{figure}

6) There may be two AGN populations: 1) a ``pure'' RQ population A
which shows little overlap with the RL domain (e.g. it includes
$\sim$70\%\ of the RQ sample) and 2) a mixed RL-RQ population B
($\sim$30\%\ of the RQ sources and almost all RL sources). We
recently summarized \citep{Sulentic07} the empirical (and inferred
theoretical differences) between pop. A and B sources which, in
fact, encompass almost all existing multiwavelength measures of AGN.
A recent comparison between FWHM and line dispersion (2nd moment of
the profile) measurements for H$\beta$ has accidentally confirmed
our two population hypothesis \citep{Collin06}. Focussing on the key
4DE1 parameters we can say that Pop. A sources show relatively
narrow single-component Balmer lines, strong FeII emission, a CIV
blueshift and a soft X-ray excess. Pop. B sources show broader, and
often more complex, Balmer lines, weak FeII emission and absence of
a CIV blueshift or soft X-ray excess.

Average Balmer line spectra in the 4DE1 context \citep{Sulentic02}
require a minimum of two Gaussian components for a reasonable model
fit. Aside from the relatively unshifted component assumed to be the
``classical'' BLR we find a broader and redshifted (VBLR- Very Broad
Line Region) component. The latter component is dangerous for
reverberation studies and BH mass estimation because it responds to
continuum changes differently than the classical BLR
\citep{Sulentic00c}. At the upper extreme of optical 4DE1 space we
find a very small number ($<$1\%) of sources showing double-peaked
Balmer line profiles. While they have been interpreted a a direct
signature of accretion disk line emission \citep{EH94} there are
many problems with this hypothesis \citep{Sulentic99,Sulentic06}.
The situation is reversed for UV high ionization line profiles like
CIV1549. Pop. A sources require unshifted and blueshifted components
while pop. B sources show stronger and less shifted line profiles.
Extreme pop. A sources are sometimes dominated by the blueshifted
component.

CIV measures have been clouded in controversy for many years with
much of the disagreement centered on the reality and strength of a
narrow CIV emission component \citep{SM99}. We recently presented a
recipe for reasonable CIV narrow component subtraction
\citep{Sulentic07} and show that corrected broad line measures
clarify our picture of high ionization line emission properties in
AGN.

7) Figure 2 shows the current best CIV-\gs\ 4DE1 plane which
illustrates two of the pop. A-B differences summarized above. Figure
2 suggests that pop. B sources show a much stronger domain space
concentration than pop. A. \rfe\ and CIV shift (c(1/2)) measures
also show this concentration. While pop. A sources show considerable
scatter in all 4DE1 measures, pop B sources show this scatter only
in measures of FWHM H$\beta$. This may be telling us that the
physics/geometry/kinematics of pop. B sources is much more similar
from source-to-source. This makes some sense if we view pop. B
sources as the end phase of quasar evolution when the accretion rate
is low.

X-ray measures shown in Figure 2 are based largely on archival ROSAT
observations. We have only begun to harvest the new generation of
XMM-Newton X-ray measures. Three recent papers have analyzed
strongly overlapping samples of PG quasars ranging from 20-40
sources \citep{Porquet04,Bro06,Pic05}. These samples allow a
preliminary test of earlier inferences that motivated us to adopt
\gs\ as a key 4DE1 parameter. Table 1 summarizes a comparison of
mean XMM measures for PG quasars observed so far. We restrict this
preview to low z PG quasars (included in BG92) for the following two
reasons: 1) higher redshift PG sources, assumed to have similar
intrinsic X-ray luminosity, will yield lower s/n spectra on average
and 2) we have no FWHM H$\beta$ or \rfe\ measures for higher z PG
sources precluding a contextualization in 4DE1 space. We immediately
distinguish between our so-called population A and B sources. Even
if this distinction turns out not to be fundamental in the sense of
physically discrete quasar classes it has proven effective for
highlighting differences across the 4DE1 sequence.

\begin{table}

\caption{}
\centering
\begin{tabular}{cccc}
\hline \hline
\multicolumn{4}{c}{Piconcelli--single power-law fit} \\
\hline Pop. & N & $\Gamma_{2-12keV}$ & $\Gamma_{0.3-12keV}$\\
\hline

A    & 22  &   1.94  &  2.71 \\
B    & 11  &   1.66  &  1.88 \\

\hline \hline

\multicolumn{4}{c}{Pourquet/Brocksopp/Piconcelli--broken power-law fit} \\
\hline
Pop. & N & $\Gamma_{soft}$ & $\Gamma_{hard}$ \\
\hline
A    & 14/14/13  &   2.825/2.86/2.98  &  2.07/2.13/1.39 \\
B    & 7/6/8     &   2.31/2.61/2.75   &  1.795/1.82/1.38 \\
\hline
\end{tabular}

\end{table}

The largest difference between pop. A and B sources involves
comparison over the largest energy range (0.3-12keV). The pop. A-B
soft photon index difference is larger than the one reported in the
defining papers of the 4DE1 concept \citep{Sulentic00a,Sulentic00b}.
It is also 3$\times$ larger than the A-B difference in the
``harder'' 2-12keV energy range. This was the motivation for
suggesting that only pop. A sources show a soft X-ray excess. XMM
measures confirm the ROSAT derived pop. A-B difference and Table 1
shows that the mean soft 0.3-12 keV index for pop. B sources is
similar to harder pop. A-B indices derived over the 0.3-12keV range.
Despite claims that all PG quasars show a soft excess, this result
suggests that pop. B source do not.

Single power-law fits to PG X-ray spectra are generally poor which
motivated the above studies to consider broken power-laws providing
both hard and soft photon indices as given in Table 1. We see some
evidence for a stronger pop A-B difference in the soft band however
there is considerable variance among the three studies. If we
restrict the \citealt{Bro06} sample to broken power-law fits with
the best $\chi$$^2$ solutions we find a soft component pop. A-B
difference $\Delta$=3.0-2.5$\sim$0.5 which supports a pop. A soft
excess at a K-S significance level P=0.002. Sigma values are 0.6 and
0.3 also confirming the smaller scatter among pop B sources
mentioned above. Clearly we need more X-ray spectra for low z PG
quasars.

As also reported earlier \citep{Sulentic00a,Sulentic00b} a (smaller)
difference is seen for harder X-ray measures in the \citealt{Pic05}
sample. Pop. B sources show a harder spectrum than pop. A sources in
most of the XMM studies. All three XMM studies report correlations
between FWHM H$\beta$ and soft/hard measures confirming the
correlation that we found with a larger (n=112) ROSAT sample
\citep{Sulentic00a,Sulentic00b}. The apparently continuous
correlation may be telling us that Pop. A-B are simply the ends of
single AGN sequence rather than two distinct classes.

The ``soft X-ray excess'' has been widely interpreted as the thermal
signature of an accretion disk \citep{Pounds95}. This interpretation
is reasonable within the 4DE1 context where the strongest soft
excess is found among sources with the the smallest black hole
masses and highest accretion rates (highest L/L$_{Edd}$) Following
this reasoning \citet{Pic05} explored composite model fits. It is
interesting that models employing a power-law+bremstrahlung yielded
a much larger pop. A-B difference (kT=0.02 vs. 0.07) than models
employing a power-law+black body. If the pop. A-B difference is real
then this may be telling us that the thermal disk interpretation is
incorrect or too simple. \citet{Bro06} reached a similar conclusion
using a different argument. It has been known for a long time that
the pop. A soft excess implies too high temperatures which has
motivated some to explore Comptonized disk models
\citep[e.g.][]{HMG94}.

Two conclusions/comments from the XMM studies \citep{Pic05} require
a response within the 4DE1 context. A) A RL-RQ difference could
suggest the presence of an extra continuum contribution from self
synchrotron jet emission \citep{Zamorani81}. 4DE1: We suggest that
the few RL sources in PG preclude statistically useful inferences.
However in 4DE1 context we know that RQ pop. B sources show very
similar properties to the (largely pop. B) RL sources. Our much
larger ROSAT sample shows this statement to be reasonable. This
motivated us to opine that since RQ pop. B sources show the same
spectra as RL sources there is no evidence for an X-ray component
related to a sources radio-loudness. B) None of the two component
models gives a satisfactory fit for all sources. This result
suggests that the shape of the soft excess is not a universal QSO
property. 4DE1: The population A-B distinction is very useful here
because it suggests that there are two fundamentally different kinds
of X-ray emitting quasars. Thew soft X-ray excess is only a property
of one of these populations.

8) We recently harvested from the HST archive all UV spectra (n=130
sources) covering the CIV$\lambda$1549 region. We find mean CIV
profile shifts (at FWHM) of -677 km s$^{-1}$ and -39 km s$^{-1}$
respectively for pop. A and B sources confirming the original
motivation for adopting this measure as a key 4DE1 parameter
\citep{Sulentic00b, Sulentic07}. The blueshift of high ionization UV
emission lines is therefore concentrated in sources with FWHM
H$\beta$$<$4000 km s$^{-1}$. This means that it is a largely RQ
phenomenon (mean CIV profile shifts for RQ and RL sources are -582
km s$^{-1}$ and +52 km s$^{-1}$, respectively); CIV equivalent width
also shows a striking pop. A-B difference: 117$\AA$ and 57$\AA$,
respectively. We recently discovered an apparent correlation between
CIV FWHM and EW measures, but only for pop. A sources. As mentioned
earlier most pop. B sources show CIV shift/EW, \rfe and \gs\
measures that are identical within measurement uncertainties. We
hope to exploit CIV measures as an orientation indicator using this
new correlation. It is now well known that the Baldwin effect is
dominated by intrinsic (we would say 4DE1) rather than cosmological
effects \citep{Bachev04,BL04}.

{\bf PHYSICAL INFERENCES:} We have described some of the empirical
results connected with the four key parameters defining 4DE1 space.
4DE1  is the most  effective spectroscopic unifier available, at the
same time it emphasizes the differences between AGN. We have also
begun to explore the physical implications of 4DE1
\citep{Marziani01,Marziani03b} and suggest the following theoretical
inferences from the empiricism; adopting a model where low
ionization broad lines (e.g. broad H$\beta$ and FeII emission) arise
in a photoionized medium in or near an accretion disk. Simple models
then allow us to use: a) line reverberation to infer the radius of
the broad line region \citep{Kaspi00,Kaspi05}, b) line FWHM or
dispersion to infer the BH mass (assuming a virialized medium:
\citealt{Marziani03b,Collin06,Sulentic06}) and c) \rfe\ to infer the
density of the FeII emitting medium that is as dense as or denser
than the region producing H$\beta$ \citep[][and this
volume]{Marziani01}. The Eddington ratio emerges from b) + source
bolometric luminosity. The most useful summary statement here might
be a comparison of median inferred properties for pop. A and B
sources. Table 2 summarizes the inferred pop. A-B values taken from
the following papers: \citealt{Marziani01,Marziani03b,Sulentic06}.
Figure 3 also shows some of the inferred physical trends thought to
drive source occupation in the optical plane of 4DE1.  See also the
review by P. Marziani in this volume.

\begin{table}
\caption{}
\centering
\begin{tabular}{cccc}

\hline \hline

\multicolumn{3}{c}{Inferred Physical Properties}\\

\hline

QUANTITY  &   POP--A   &   POP--B \\

\hline

log n$_{E}$      &   11.5      &  9.5  \\

log U        &     -1.5    &    -1.0 \\

log M$_{BH}$   &    7.7     &    9.0 \\

L/L$_{Edd}$    &   0.3     &   0.08 \\

\hline \hline

\end{tabular}
\end{table}

\begin{figure}
  \includegraphics[width=\columnwidth]{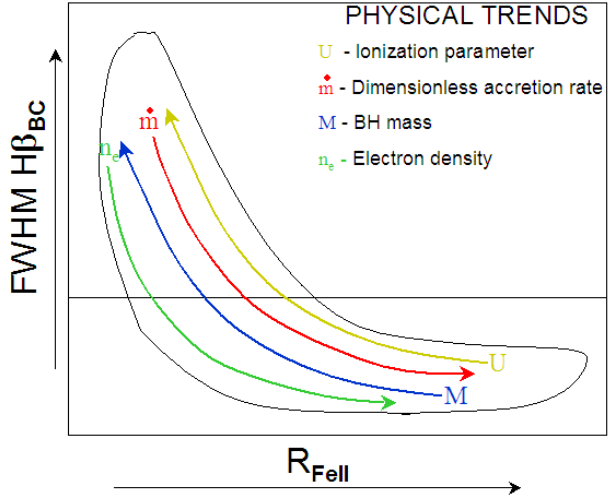}
  \caption{Inferred physical trends along the quasar sequence in the
  optical plane of 4DE1 space.}
  \label{fig:simple}
\end{figure}

\citet{Collin06} recently argued that a thin flat BLR disk structure
could be ruled out. We usually assume that H$\beta$, and especially
FeII, in high accreting pop. A sources arise in a flatter
distribution than pop. B. This assumption is motivated by
empiricism. We can either: 1) compare the typical FWHM H$\beta$
values for the most face-n pop. A and B sources or 2) the dispersion
in observed FWHM values for each population assuming that we observe
the BLR over the same range of viewing angles (e.g. 0$^\circ$ to
45$^\circ$). We also assume that H$\beta$ and FeII emission arise
from rather similar cloud distributions because FWHM FeII $\sim$
FWHM H$\beta$. The narrowest pop. A and B sources show FWHM
H$\beta$$\sim$ 800 km s$^{-1}$ and 4000 km s$^{-1}$, respectively.
These are then assumed to be the ``face-on'' values for the two
populations and are measuring motions largely perpendicular to the
disk plane where Keplerian rotation dominate. The much smaller
velocity dispersion in pop. A sources is consistent with the idea of
a highly flattened distribution. Either pop. B sources are much less
flattened or there is an additional emission component producing the
much higher observed velocity dispersion perpendicular to the disk.
A more extreme alternative would see line emission in pop. B sources
arising entirely in a non-disk configuration.

Note that the ``face-on'' value for pop. A sources is set by the
lower limit of observed FWHM H$\beta$ (with FeII emission
corroborating that the source is an AGN and that a dense enough
medium exists to account for the strong FeII emission). The
``edge-on'' value of FWHM H$\beta$ for pop. A is set by the fact
that sources with \rfe\ $>$0.4 (unambiguously Pop. A) show FWHM
H$\beta$$<$4000 km s$^{-1}$. The situation is less confused for pop.
B sources because we can infer the range of FWHM H$\beta$ from RL
sources via radio morphology and core/lobe flux ratios
\citep{Rokaki03,Sulentic03}. We then assume that pop. B RQ sources
follow RL pop. B because they are co-spatial in 4DE1. Pop. B sources
start to become rare at about FWHM H$\beta$=10$^4$km/s. If we assume
the same range of viewing angle for pop. A and pop. B sources then
we observe velocity ranges of a factor of  $\sim$5 and $\sim$2.5 for
pop. A and B respectively, consistent with a flatter pop. A emitting
cloud distribution.

The typical inferred properties in Table 2 are representative of our
current understanding of sources with z$\leq$0.8. Unfortunately the
H$\beta$ spectral region cannot be studied from the ground beyond
z$\sim$1.0 (z=0.7 for SDSS spectra) without major effort. Ground
based CIV measures have been extensively used to extract M$_{BH}$
and other properties out to much higher redshift (z$\sim$4.0). Our
own studies of the CIV line using the HST archive suggest that CIV
cannot be trusted for M$_{BH}$ estimation for many reasons
\citep{Sulentic07,Netzer07}: 1) the obvious Malmquist bias
associated with CIV measures for any optically selected sample, 2)
the complexity of the CIV profile and properties reflected in pop.
A-B profile differences and the lack of a clear correlation between
FWHM H$\beta$ -- FWHM (CIV) and 3) especially for the pop. A
(majority) of RQ sources - the blueshifted and blue asymmetric
profile shapes which must raise doubts that the high ionization UV
lines arise in a virialized medium. Sources with a strong narrow
line component (likely more common among pop. B sources if lines
like [OIII]$\lambda$5007 are any guide) tend to give a false sense
of security because they make CIV profiles look more symmetric and
less blueshifted. Recent observations of high z type 2 AGN reveal
frequent and reasonable strong CIV narrow line emission (see
references in \citealt{Sulentic07}).

We have chosen to make a major effort with VLT ISAAC and now have IR
spectra of the H$\beta$ region for 50+ sources between z=0.8-2.5.
VLT enables us to obtain IR spectra with resolution and s/n similar
to ground based spectra for lower redshift sources.  Results so far
suggest that source occupation in 4DE1 space is similar up to the B=
-29 to -30 source luminosity range. Much remains to be done but 4DE1
space continues to offer a useful way to contextualize both
empirical and physical properties of quasars. It offers the hope of
simplifying or resolving many of the current questions about these
most active objects.

We thank the organizers for the opportunity to speak in Huatlulco
where we celebrated the birthday of Deborah Dultzin who has been a
member of our collaboration for 10+ years.

\end{document}